\documentclass{JHEP3}
\let\ifpdf\relax
\usepackage{ifpdf}
\usepackage{multicol,bbm}
\usepackage{graphicx,psfrag}
\usepackage{dcolumn}
\usepackage{bm}
\usepackage{amsmath,amssymb, bbold}
\usepackage{color}
\let\normalcolor\relax

\newcommand{\be}{\begin{equation}}
\newcommand{\ee}{\end{equation}}
\newcommand{\bea}{\begin{eqnarray}}
\newcommand{\eea}{\end{eqnarray}}

\newcommand{\eq}{\begin{equation}}
\newcommand{\eqe}{\end{equation}}

\newcommand{\eqa}{\begin{eqnarray}}
\newcommand{\eqae}{\end{eqnarray}}

\title{Bimetric gravity doubly coupled to matter: theory and cosmological implications}

\author{Yashar Akrami, Tomi S. Koivisto, David F. Mota and Marit Sandstad\\
       Institute of Theoretical Astrophysics, University of Oslo\\
       P.O. Box 1029 Blindern, N-0315 Oslo, Norway\\
       E-mail: \email{yashar.akrami@astro.uio.no, t.s.koivisto@astro.uio.no, d.f.mota@astro.uio.no, marit.sandstad@astro.uio.no}}

\received{\today}
\accepted{}

\abstract{A ghost-free theory of gravity with two dynamical metrics both coupled to matter is shown to be consistent and viable. Its cosmological implications are studied, and the models, in particular in the context of partially massless gravity, are found to explain the cosmic acceleration without resorting to dark energy.}

\preprint{}

\begin{document}

\hyphenation{}

\section{Introduction}

Theories of interacting symmetric two-tensors~\cite{Isham:1971gm} have found a nonlinear completion~\cite{Hassan_et_al2012a,Hassan_et_Rosen2012a,Hassan:2011ea}. As one of the spin-2 particles necessarily acquires a mass~\cite{Boulanger:2000rq}, the obstacle had been the same as in massive gravity~\cite{Fierz_et_Pauli1939}: {\it a ghost mode that is excited in curved backgrounds}~\cite{Boulware_et_Deser1939}. However, that mode is avoided when the mutual interactions of the two tensor fields are composed from the recently discovered unique set of terms~\cite{deRham_et_Gabadadze2010b,deRham_Gabadadze_et_Tolley2010,deRham_Gabadadze_et_Tolley2011b,deRham_Gabadadze_et_Tolley2011c,Hassan_et_Rosen2011a,Hassan:2011hr}. The two tensors appear in these interaction terms completely symmetrically. One may then wonder whether they could also interact with other fields, i.e. the matter sector, in a symmetric way. Indeed, arguably by proper semantics, in a {\it bimetric theory of gravity} both of the metrics should couple directly to matter, but as far as we know, such a ``double coupling" has not been introduced previously. That is what we shall pursue in this paper.  

We will assume, as conventional and simple, that the dynamics for both of the metrics are given by the corresponding Einstein-Hilbert terms.\footnote{More general terms could be considered too~\cite{Nojiri:2012zu,Nojiri:2012re}, without changing our conclusions essentially, though the cosmological dynamics would then be affected. On other extensions, see e.g.~\cite{Paulos_et_Tolley2012,Cai:2012ag,Huang:2012pe,Wu:2013ii,Leon:2013qh}.} The version of massive gravity~\cite{Hinterbichler2011} where the ``background metric'' is not given any dynamics but kept fixed has been studied actively and found to suffer from several problems, in particular the absence of consistent simplest solutions corresponding to cosmological~\cite{D'Amico_et_al2011,Comelli_et_al2011b,De_Felice_Gumrukcuoglu_etMukohyama2012,Tasinato:2012ze,Koyama:2011wx,Comelli:2012db,DeFelice:2013awa,DeFelice:2013bxa,Khosravi:2013axa} or  black hole space-times~\cite{Berezhiani:2011mt,Babichev:2013una,Berezhiani:2013dw}, and the theory has been claimed to inhabit tachyons~\cite{Chamseddine:2013lid,Deser:2012qx}. The dynamical bimetric version, which we will henceforth refer to as the Hassan-Rosen theory, however, seems to avoid these problems (see though \cite{Deser:2013gpa}), and for example admits consistent cosmological solutions~\cite{Volkov2012a,vonStrauss:2011mq,Khosravi:2012rk,Berg:2012kn} which furthermore produce the observed late-time acceleration without introducing an explicit cosmological constant~\cite{Volkov:2012cf,Volkov:2012zb,Akrami:2012vf,Akrami:2013pna}.\footnote{We emphasise here that even though it has been shown that the version of the ghost-free bigravity theory we study here can provide a self-acceleration mechanism, it does not solve the old cosmological constant problem, namely why the observed value of the cosmological constant is orders of magnitude smaller than the value predicted by the standard model of particle physics (see e.g.~\cite{Martin:2012bt} for a recent review). Here we assume that this problem will eventually be ``naturally'' explained by other proposals, for example a successful implementation of the degravitation mechanism~\cite{Dvali:2007kt} (for possible links between theories of massive gravity and degravitation through non-local generalisations, see e.g.~\cite{Modesto:2013jea}). Similar to dark energy or many other modified-gravity explanations of the cosmic acceleration, we assume here that some yet-to-be-discovered mechanism makes the vacuum energy vanish or get screened, and the acceleration is caused by a different mechanism which in our case is massive gravitons. We should note further that the effects of quantum corrections on the mass terms and the scale(s) introduced in the theory are not yet entirely understood and whether a specific scale is technically natural (contrary to the cosmological constant) needs to be investigated (see~\cite{deRham:2013qqa} for a recent work on this issue).} In addition, spherically symmetric solutions~\cite{Comelli_et_al2011a} and black holes~\cite{Baccetti:2012ge,Sakakihara:2012iq,Volkov:2012wp,Volkov:2013roa,Brito:2013wya}, as well as the limits to the fixed-metric massive gravity~\cite{Baccetti:2012bk} and to the partially massless gravity~\cite{Hassan:2012gz,Hassan:2012rq,Hassan:2013pca} have been studied. The absence of ghosts has been discussed from several points of view~\cite{Golovnev:2011aa,Hassan:2011ea,Hassan:2012qv,Kluson:2012wf,Nomura:2012xr,Golovnev:2013fj,Kluson:2013cy}, and the theory has been generalised to higher numbers of interacting spin-2 particles~\cite{Hinterbichler_et_Rosen2012,Hassan:2012wt,Hassan:2012wr}.  

Here we show that matter can be consistently coupled to the two metrics in a generalisation of the ghost-free Hassan-Rosen theory. If the potential interactions vanish, the theory reduces to two copies of pure Einstein gravity, which can {\it not} both be coupled to matter. We should also clarify that we do not assume the existence of a new kind of hidden matter that would couple solely to the ``background" metric. Such separate matter sectors have been considered previously and shown to lead to violations of energy conditions~\cite{Baccetti:2012bk,Baccetti:2012re} and degeneracies in interpreting the observed phenomena~\cite{Capozziello:2012re}. We instead assume only ``usual" matter but allow it to couple to the other metric as well, and this results in a possibility to efficiently constrain the coupling with local gravity experiments. In the framework of bimetric variational principle, where an independent metric generates the geometric and affine structures of space-time~\cite{Amendola:2010bk,Koivisto:2011vq,BeltranJimenez:2012sz} (see also~\cite{Hossenfelder:2008bg}), the connection and the metric in the matter action are independent fields and thus spinning matter naturally couples to two different metrics. However, these set-ups are again different from the case we study here where the identical metric sectors have identical forms of the coupling to the matter sector. Here, we cannot fix the coupling constants from first principles and they remain free parameters of the theory to be constrained by observations.  

The structure of the paper is as follows. In section~\ref{tdcbm} we present the ghost-free bimetric theory of gravity with the coupling of the background metric to matter and show that it is consistent in view of the Bianchi identities~\cite{Koivisto:2005yk}. We also illustrate the system concretely by considering the Lagrangians of a scalar field and a point particle. In section~\ref{thaib} we then, as a relevant example, apply the resulting theory to the cosmologically interesting case of an isotropic and homogeneous space-time, and investigate in particular how the introduction of the new coupling changes the dynamics of the background. We then in section~\ref{luca} particularise our discussions of the cosmological implications of the theory to the late-time cosmic evolution where relativistic particles contribute to the energy budget of the universe only negligibly. We show how the set of dynamical equations can be simplified in this dust-dominated universe and how the theory can give rise to an exact $\Lambda$CDM regime with an emergent cosmological constant. We finally, in section~\ref{spr}, demonstrate this further, by studying various interesting subsets of the full parameter space, and discuss how a late-time acceleration consistent with cosmological observations can be achieved in our bimetric theory, just as in the standard $\Lambda$CDM model of cosmology. Section~\ref{c} then briefly summarises our findings.

\section{The doubly-coupled bimetric gravity}
\label{tdcbm}

\subsection{Theory}

The original formulation of the Hassan-Rosen theory was first given in \cite{Hassan_et_Rosen2012a}, where the action of the theory takes the form:
\begin{eqnarray}
  S & = & - \frac{M^2_g}{2}\int d^4x\sqrt{-\det g}R(g) \nonumber \\
  && - \frac{M^2_f}{2}\int d^4x\sqrt{-\det f}R(f) \nonumber\\
  && + m^2M_g^2\int d^4x\sqrt{-\det g}\sum_{n=0}^{4}\beta_ne_n\left(\sqrt{g^{-1}f}\right) \nonumber \\
  && + \int d^4x\sqrt{-\det g}\mathcal{L}_m\left(g, \Phi\right).\label{eq:ActionOriginal} 
\end{eqnarray}

Here, $f$ and $g$ denote two 2-tensors of the gravity sector, with metric properties, whose dynamics are given by two corresponding Einstein-Hilbert terms, as well as five interaction terms (third line in the action). Defining $\mathbbm{X}\equiv\sqrt{g^{-1}f}$, $e_n(\mathbbm{X})$ are polynomials of the eigenvalues of the matrix $\mathbbm{X}$:
\begin{eqnarray}
e_0\left(\mathbbm{X}\right) & \equiv & 1, \qquad \nonumber \\
e_1\left(\mathbbm{X}\right) & \equiv & \left[\mathbbm{X}\right], \qquad \nonumber \\
e_2\left(\mathbbm{X}\right) & \equiv & \frac{1}{2}\left(\left[\mathbbm{X}\right]^2 - \left[\mathbbm{X}^2\right]\right),\nonumber \\
e_3\left(\mathbbm{X}\right) & \equiv & \frac{1}{6}\left(\left[\mathbbm{X}\right]^ 3 - 3\left[\mathbbm{X}\right]\left[\mathbbm{X}^2\right] + 2\left[\mathbbm{X}^3\right]\right), \qquad \nonumber \\
e_4\left(\mathbbm{X}\right) & \equiv & \det\left(\mathbbm{X}\right),
\end{eqnarray}
where square brackets denote traces of the matrices. The quantities $\beta_n (n=0,...,4)$ and $m$ are free parameters,\footnote{Obviously $m$ can be absorbed into the parameters $\beta_n$ and is therefore {\it not} an independent free parameter of the theory. Writing $m$ as a separate parameter here is more of a convention that we adhere to.} and $M_g$ and $M_f$ denote two (Planck-like) mass scales corresponding to $g$ and $f$, respectively. The last term in eq. (\ref{eq:ActionOriginal}) gives the interaction between $g$ (which is assumed to be the ``physical metric") and matter, where $\mathcal{L}_m\left(g, \Phi\right)$ denotes the matter Lagrangian (corresponding to e.g. the standard model of particle physics), with matter fields $\Phi$.

Strictly speaking, the theory in this original formulation is not a ``bimetric" theory of gravity because only one of the two tensor fields, namely $g$, directly couples to matter; this is therefore rather a theory of ``gravity coupled to matter and a symmetric 2-tensor"~\cite{Berg:2012kn}.

As a natural, next step in generalising the theory, we therefore investigate the properties and implications of the case where both tensor fields couple to matter, and to keep the theory as symmetric and simple as possible, we let $f$ couple to matter in the same way as $g$ does, i.e. ``minimally". The action (\ref{eq:ActionOriginal}) now becomes:
\begin{eqnarray}
  S &=& - \frac{M^2_g}{2}\int d^4x\sqrt{-\det g}R(g) \nonumber \\
  && - \frac{M^2_f}{2}\int d^4x\sqrt{-\det f}R(f) \nonumber\\
  && + m^2M_g^2\int d^4x\sqrt{-\det g}\sum_{n=0}^{4}\beta_ne_n\left(\sqrt{g^{-1}f}\right) \nonumber\\
  && + \int d^4x\sqrt{-\det g}\mathcal{L}_m\left(g, \Phi\right) \nonumber \\
  && + \alpha\int d^4x\sqrt{-\det f}\mathcal{L}_m\left(f, \Phi\right),\label{eq:ActionDC} 
\end{eqnarray}
where $\alpha$ parameterises the relative couplings of matter to the metrics $g$ and $f$.

The equations of motion for the two metrics (or the double set of Einstein field equations) now read:
\footnotesize{
\begin{eqnarray}
R_{\mu\nu}^g - \frac{1}{2}g_{\mu\nu} R^g + \frac{m^2}{2}\sum_{n=0}^3\left(-1\right)^{n}\beta_n\left[g_{\mu\lambda}Y^{\lambda}_{(n)\nu}\left(\sqrt{g^{-1}f}\right) + g_{\nu\lambda}Y^{\lambda}_{(n)\mu}\left(\sqrt{g^{-1} f}\right)\right] = \frac{1}{M_g^2}T^g_{\mu\nu}, \label{eq:Einsteingeng}\\
R_{\mu\nu}^f - \frac{1}{2}f_{\mu\nu} R^f + \frac{m^2}{2M^2_\star}\sum_{n=0}^3\left(-1\right)^{n}\beta_{4 - n}\left[f_{\mu\lambda}Y^{\lambda}_{(n)\nu}\left(\sqrt{f^{-1} g}\right) + f_{\nu\lambda}Y^{\lambda}_{(n)\mu}\left(\sqrt{f^{-1} g}\right)\right] = \frac{\alpha}{M_f^2}T^f_{\mu\nu}, \label{eq:Einsteingenf}
\end{eqnarray}}\normalsize
where $R_{\mu\nu}^g$ and $R_{\mu\nu}^f$ are Ricci tensors, $R^g$ and $R^f$ are Ricci scalars, $T^g_{\mu\nu}$ and $T^f_{\mu\nu}$ are stress-energy tensors corresponding to the metrics $g$ and $f$, respectively, and $M_\star^2 \equiv M_f^2/M_g^2$. The functions $Y_{(n)}\left(\mathbbm{X}\right)$ take the following forms:
\begin{eqnarray}
  Y_{(0)}\left(\mathbbm{X}\right) &\equiv& \mathbbm{1}, \quad \nonumber \\
  Y_{(1)}\left(\mathbbm{X}\right) &\equiv& \mathbbm{X} - \mathbbm{1}\left[\mathbbm{X}\right],\quad \nonumber \\
  Y_{(2)}\left(\mathbbm{X}\right) &\equiv& \mathbbm{X}^2 - \mathbbm{X}\left[\mathbbm{X}\right] + \frac{1}{2}\mathbbm{1}\left(\left[\mathbbm{X}\right]^2 - \left[\mathbbm{X}^2\right]\right), \nonumber \\  
  Y_{(3)}\left(\mathbbm{X}\right) &\equiv& \mathbbm{X}^3 - \mathbbm{X}^2\left[\mathbbm{X}\right] + \frac{1}{2}\mathbbm{X}\left(\left[\mathbbm{X}\right]^2 - \left[\mathbbm{X}^2\right]\right), \nonumber \\
  && - \frac{1}{6}\mathbbm{1}\left(\left[\mathbbm{X}\right]^ 3 - 3\left[\mathbbm{X}\right]\left[\mathbbm{X}^2\right] + 2\left[\mathbbm{X}^3\right]\right).
\end{eqnarray}

As noted in \cite{Berg:2012kn}, and emphasised in~\cite{Akrami:2012vf}, in the singly-coupled bigravity the mass ratio $M_\star$ can always be set to unity by a suitable rescaling of $f_{\mu\nu}$ and the parameters $\beta_n$. In our doubly-coupled generalisation, the same scaling can be performed, but in this case the rescaling will have to extend also to $\alpha$ and all coupling constants in the $f$-coupled matter Lagrangian $\mathcal{L}_m\left(f, \Phi\right)$. This will in general lead to a complication rather than simplification of the theory, and we therefore leave $M_\star$ as a free parameter except for specific cases of interest in cosmology as we will discuss further below.

\subsection{Bianchi constraints}

Combining Bianchi identities of the two metrics $g$ and $f$ with the field equations (\ref{eq:Einsteingeng}) and (\ref{eq:Einsteingenf}) gives rise to the following Bianchi constraints that couple the $f-g$ interaction terms to the two stress-energy tensors $T^g_{\mu\nu}$ and $T^f_{\mu\nu}$:
\small{
\begin{eqnarray}
 \nabla_g^{\mu}\frac{m^2}{2}\sum_{n=0}^3\left(-1\right)^{n}\beta_n\left[g_{\mu\lambda}Y^{\lambda}_{(n)\nu}\left(\sqrt{g^{-1}f}\right) + g_{\nu\lambda}Y^{\lambda}_{(n)\mu}\left(\sqrt{g^{-1} f}\right)\right]= \frac{1}{M_g^2} \nabla_g^{\mu}T^g_{\mu\nu}, \label{eq:Bianchig}\\
 \nabla_f^{\mu}\frac{m^2}{2M^2_\star}\sum_{n=0}^3\left(-1\right)^{n}\beta_{4 - n}\left[f_{\mu\lambda}Y^{\lambda}_{(n)\nu}\left(\sqrt{f^{-1} g}\right) + f_{\nu\lambda}Y^{\lambda}_{(n)\mu}\left(\sqrt{f^{-1} g}\right)\right]= \frac{\alpha}{M_f^2}\nabla_f^{\mu}T^f_{\mu\nu}. \label{eq:Bianchif}
\end{eqnarray}}\normalsize

The conservation equation resulting from the general covariance of the complete matter sector now reads:
\begin{equation}\label{eq:ParticleEOM}
  \sqrt{-\det g}\nabla_g^\mu T^g_{\mu\nu} + \alpha\sqrt{-\det f}\nabla_f^\mu T^f_{\mu\nu} = 0.
\end{equation}

Combining this with the Bianchi constraints (\ref{eq:Bianchig}) and (\ref{eq:Bianchif}) imposes the following constraint on the interactions of the metrics:
\footnotesize{
\begin{eqnarray}
 &&\sqrt{-\det g}\nabla_g^{\mu}\frac{M_g^2m^2}{2}\sum_{n=0}^3\left(-1\right)^{n}\beta_n\left[g_{\mu\lambda}Y^{\lambda}_{(n)\nu}\left(\sqrt{g^{-1}f}\right) + g_{\nu\lambda}Y^{\lambda}_{(n)\mu}\left(\sqrt{g^{-1} f}\right)\right] \nonumber\\
 +&&\sqrt{-\det f}\nabla_f^{\mu}\frac{m^2M_g^2}{2}\sum_{n=0}^3\left(-1\right)^{n}\beta_{4 - n}\left[f_{\mu\lambda}Y^{\lambda}_{(n)\nu}\left(\sqrt{f^{-1} g}\right) + f_{\nu\lambda}Y^{\lambda}_{(n)\mu}\left(\sqrt{f^{-1} g}\right)\right]~=~0.\label{eq:BianchiCombined}
\end{eqnarray}}\normalsize

This equation can be derived directly from the general covariance of the bimetric mass (or metrics' interaction) terms in the action (\ref{eq:ActionDC}) and always holds regardless of the matter coupling. In addition, this expression is an identity that holds for any choices of the metrics $g$ and $f$~\cite{HassanPrivate}.

For the original Hassan-Rosen theory, where only one metric (namely $g$) couples to matter, eq. (\ref{eq:ParticleEOM}) implies that the corresponding stress-energy tensor $T^g_{\mu\nu}$ is conserved individually. It is tempting to impose the same conservation to the stress-energy tensors $T^g_{\mu\nu}$ and $T^f_{\mu\nu}$ separately also for the case of the doubly-coupled bimetric gravity. This would then give two conditions on the metrics that relate them through the equations (\ref{eq:Bianchig}) and (\ref{eq:Bianchif}).\footnote{Since the two expressions on the right hand sides of eqs. (\ref{eq:Bianchig}) and (\ref{eq:Bianchif}) are closely related, there would likely be only one constraint from the two Bianchi constraints. This is for instance the case for a homogenous and isotropic background, as shown in~\cite{vonStrauss:2011mq}.} However, this may be a dangerous route to follow which can rule out physically viable models. For instance, there could be an energy transfer between the couplings of the two metrics and the normal energy content of the universe, or this transfer could lead to an overall transfer of energy from the stress-energy tensor seen by $g$ to the one seen by $f$, or vice versa. We therefore do not impose such non-trivial conditions and instead use the full Bianchi constraints (\ref{eq:Bianchig}), (\ref{eq:Bianchif}) and (\ref{eq:BianchiCombined}).

\subsection{Matter equations of motion and differences to two copies of GR}

The doubly-coupled bimetric theory that we propose here is fundamentally different from a theory in which two {\it non-interacting} 2-tensors couple to matter minimally, i.e. a theory that resembles two identical copies of General Relativity (GR). In order to demonstrate this, we consider the simple example of a scalar field $\phi$, as the only matter component, that couples to both metrics $g$ and $f$ according to the action (\ref{eq:ActionDC}), with matter Lagrangian densities (most of the discussion here is based on~\cite{HassanPrivate}):
 \begin{eqnarray}
  \mathcal{L}_m^g &\equiv& \mathcal{L}_m\left(g, \phi\right)~=~\frac{1}{2}g^{\mu\nu}\partial_\mu\phi\partial_\nu\phi-V(\phi), \label{eq:scalarLagrangg}\\
  \mathcal{L}_m^f &\equiv& \mathcal{L}_m\left(f, \phi\right)~=~\frac{1}{2}f^{\mu\nu}\partial_\mu\phi\partial_\nu\phi-V(\phi).\label{eq:scalarLagrangf}
\end{eqnarray}

The $\phi$-equation of motion, as a result of varying the full action (\ref{eq:ActionDC}) with respect to $\phi$, becomes:
\begin{eqnarray}\label{eq:phiEOM}
\frac{\delta(\sqrt{-\det g}\mathcal{L}_m^g)}{\delta\phi} + \alpha \frac{\delta(\sqrt{-\det f}\mathcal{L}_m^f)}{\delta\phi}&=& 0,
\end{eqnarray}
which together with eqs. (\ref{eq:scalarLagrangg}) and (\ref{eq:scalarLagrangf}) gives:
\begin{equation}
  \left[\Box_g + \alpha\left(\frac{\sqrt{-\det f}}{\sqrt{-\det g}}\right)\Box_f\right]\phi - \left[1 + \alpha\left(\frac{\sqrt{-\det f}}{\sqrt{-\det g}}\right)\right]V'(\phi) = 0, \label{eq:phiEOM2}
\end{equation}
where $\Box_g$ and $\Box_f$ denote the d'Alembertian operators in the $g$ and $f$ metrics, respectively, and $'$ denotes derivative with respect to $\phi$.

On the other hand, general covariance of the $g-\phi$ and $f-\phi$ interactions separately implies:
\begin{eqnarray}
\sqrt{-\det g}\nabla^{\mu}_gT^g_{\mu\nu} - \frac{\delta(\sqrt{-\det g}\mathcal{L}_m^g)}{\delta\phi}\partial_\nu\phi&=& 0, \label{eq:gencov1}\\
\sqrt{-\det f}\nabla^{\mu}_fT^f_{\mu\nu} - \frac{\delta(\sqrt{-\det f}\mathcal{L}_m^f)}{\delta\phi}\partial_\nu\phi&=& 0, \label{eq:gencov2}
\end{eqnarray}
where
\begin{eqnarray}
T^g_{\mu\nu} \equiv - \frac{2}{\sqrt{-\det g}}\frac{\delta(\sqrt{-\det g}\mathcal{L}_m^g)}{\delta g^{\mu\nu}}, \qquad T^f_{\mu\nu} \equiv - \frac{2}{\sqrt{-\det f}}\frac{\delta(\sqrt{-\det f}\mathcal{L}_m^f)}{\delta f^{\mu\nu}}.
\end{eqnarray}

These two conditions together with the full conservation equation (\ref{eq:ParticleEOM}) then give:
\begin{eqnarray}
\left[\frac{\delta(\sqrt{-\det g}\mathcal{L}_m^g)}{\delta\phi}+\alpha \frac{\delta(\sqrt{-\det f}\mathcal{L}_m^f)}{\delta\phi}\right]\partial_\nu\phi&=& 0,
\end{eqnarray}
which does {\it not} result in any additional conditions to what has already been placed on the dynamics of $\phi$ by the $\phi$-equation of motion (\ref{eq:phiEOM}).

Let us now consider the case where the two metrics do not interact directly, i.e. for $m=0$ in the action (\ref{eq:ActionDC}). The theory in this case is equivalent to coupling the same scalar field to two copies of GR. The equation of motion for $\phi$ does not change and is still given by eq. (\ref{eq:phiEOM}). Eqs. (\ref{eq:gencov1}) and (\ref{eq:gencov2}) also remain unchanged. However, the Bianchi constraints (\ref{eq:Bianchig}) and (\ref{eq:Bianchif}) in this case imply that the two stress-energy tensors $T^g_{\mu\nu}$ and $T^f_{\mu\nu}$ must be individually conserved:
\begin{eqnarray}
\nabla_g^{\mu}T^g_{\mu\nu} = 0, \qquad \nabla_f^{\mu}T^f_{\mu\nu} = 0,
\end{eqnarray}
which together with eqs. (\ref{eq:gencov1}) and (\ref{eq:gencov2}) then imply that:
\begin{eqnarray}
\frac{\delta(\sqrt{-\det g}\mathcal{L}_m^g)}{\delta\phi}\partial_\nu\phi = 0, \qquad \frac{\delta(\sqrt{-\det f}\mathcal{L}_m^f)}{\delta\phi}\partial_\nu\phi= 0.
\end{eqnarray}

These two conditions are significantly more constraining than the $\phi$-equation of motion (\ref{eq:phiEOM}) and generically imply $\phi=constant$. This shows that contrary to the case of interacting spin-2 fields, where coupling the scalar field to both metrics does not overly constrain the system, the scalar field {\it cannot} be coupled to two non-interacting spin-2 fields to result in a non-trivial theory. The manipulations here are general, therefore the conclusions hold for other fields as well as for point particles~\cite{HassanPrivate}.

For massive point particles, where the matter actions corresponding to the metrics $g$ and $f$ read:
\begin{eqnarray}
S_p^g = -m_p\int dt\sqrt{g_{\mu\nu}\dot{x}^\mu\dot{x}^\nu}, \qquad S_p^f = -m_p\int dt\sqrt{f_{\mu\nu}\dot{x}^\mu\dot{x}^\nu},
\end{eqnarray}
the geodesic equation\footnote{Strictly speaking, this equation should not be called geodesic equation, because the space-time is now equipped with two metrics and this equation does not correspond to the geodesic equation of any of the two metrics. The equation is rather just the equation of motion for the point particles which resembles a combination of the geodesic equations of the two metrics.} receives contributions from both metric terms:
\begin{equation}
  \frac{du_g^\alpha}{ds_g} + \Gamma^\alpha_{\mu\nu}u_g^\mu u_g^\nu = \alpha g^{\beta \alpha}f_{\beta \gamma}\frac{\left(g_{\mu\nu}\dot{x}^\mu\dot{x}^\nu\right)^{\frac{1}{2}}}{\left(f_{\mu\nu}\dot{x}^\mu\dot{x}^\nu\right)^{\frac{1}{2}}}\left[\frac{du_f^\gamma}{ds_f} + \bar{\Gamma}^\gamma_{\mu\nu}u_f^\mu u_f^\nu\right], \label{eq:particlegeodesic}
\end{equation}
where $ds_g = \left(g_{\mu\nu}dx^\mu dx^\nu\right)^{\frac{1}{2}}$, $ds_f = \left(f_{\mu\nu}dx^\mu dx^\nu\right)^{\frac{1}{2}}$, $u_g^\mu = dx^\mu/ds_g$ and $u_f^\mu = dx^\mu/ds_f$. $\Gamma^\alpha_{\mu\nu}$ and $\bar{\Gamma}^\alpha_{\mu\nu}$ are the Christoffel symbols in the $g$ and $f$ metrics, respectively. 

The case for massless particles is more complicated. In standard, single-metric gravity, massless particles are those that follow null geodesics of the metric, i.e. paths with $ds=0$. This means that in this case $s$ cannot be used as the affine parameter and therefore writing the geodesic equation for massless particles is not as straightforward as for massive ones. This in general results in an ambiguity in defining massless particles in the presence of two metrics because $ds=0$ for one metric {\it does not} in general imply the same for the other metric. In a particular case however, a massless particle can still be defined in an unambiguous way, i.e. when $ds$ for both metrics can be set to zero simultaneously. In this case, massless particles are those which follow the null geodesics of both metrics. Obviously this can be achieved if the two metrics are conformally related; as we will see, this case is of particular interest in various scenarios studied in the following sections on applications of the bimetric theory to cosmology.

From eq. (\ref{eq:particlegeodesic}) we see that the weak equivalence principle is obeyed by the doubly-coupled theory, although the exact trajectories followed by the point particles will deviate from the GR case. The structure involved in the derivation of this equation and equations of motion for different particle species, however, leads us to conclude that the Einstein and strong equivalence principles are violated. Constraints on the possible violations will then place constraints upon the parameter $\alpha$ as making $\alpha$ sufficiently small will make the coupling between the matter sector and the metric $f$ arbitrarily weak, and hence also make the equivalence principle violation arbitrarily small. The exact constraints from this may vary with particle species involved and details of the theory. Whether the constraints will be compatible with non-negligible extra dynamics is a question beyond the scope of this article.

\subsection{On an ambiguity in defining a physical metric}
\label{physical}

On general implications of our doubly-coupled, bimetric compared to the single-metric gravity, we must emphasise the effect of the second metric on the equations of motion of different matter components. As we observe from both eqs. (\ref{eq:particlegeodesic}) and (\ref{eq:phiEOM2}) for (massive) point particles and scalar fields, the dynamics and evolutions of matter are determined by both metrics $g$ and $f$. This means that in order to accurately know how particular particles or fields evolve, one in general {\it cannot} assume, contrary to the original Hassan-Rosen bigravity theory where only one metric directly couples to matter, that only one metric is ``physical" in the sense of being minimally coupled to matter; in this case the full geodesic or evolution equations must be considered.

To connect physical observables like redshifts of photons or universe expansion rates to the fundamental theory one has to proceed with caution. Since there is no longer a direct bridge between the geodesics of one particular metric and the trajectories of particles in general and specifically photons, there is no ``canonical'' choice of a metric, for instance $g$, from which to read off the observables, like $z = a_0/a-1$ where $a$ is the scale factor of the metric $g$. If there was a combination of the two metrics to which all matter would be minimally coupled, that would obviously be the natural choice for the physical metric instead of either $f$ or $g$. However, due to the different functional forms of couplings of different matter species to the metrics, it is obvious that no such universally minimally coupled metric exists:\footnote{One specific choice one could think of would be the massless combination of the metrics. Between $f$ and $g$ (neither of which is massless), there is no fundamental difference, since they are in a symmetric position in the original action, up to constant redefinitions of parameters.} this is a concrete manifestation of us having a theory that is not ``metric" but ``bimetric" proper. 

An apparently profound question is now, whether the predictions of the theory are unique without specifying ``by hand'' a metric among all the possible combinations of the two spin-2 tensors, that describes directly the physical space-time geometry. 
Then one should go back to the definitions of the action and the equations of motion for the particles or fields in the specific experiment and from first principles compute the observables of the theory and their connections to $f$ and $g$. Whether or how that could be done in practice for, for instance, redshifts of photons is not clear to us at this stage. It is an important point to address in future work, but in fact for all practical purposes of the present study this issue is avoided as explained below. Let us, however, note that the ambiguity here reminds us of that of the ``physical frame'' in conformally related theories in the context of Brans-Dicke theories, though here the issue is complicated by the facts that in general the relations are disformal instead of simply conformal, and that by a single transformation one cannot in general render the matter sector to be minimally coupled or the gravity sector to be pure GR.\footnote{In the (generalised) Brans-Dicke context this is not either possible if the matter couplings are not universal.}

One notes that in practice the difficulty in mapping between theory and observables disappears when the two metrics $g$ and $f$ are proportional by a constant factor, because the full equations of motion in those cases will boil down to the equations which resemble the single-metric case. This can again be seen using eq. (\ref{eq:particlegeodesic}), where setting $f_{\mu\nu}$ proportional to $g_{\mu\nu}$ reduces the geodesic equation for massive particles to a parameter-dependent, linear combination of two copies of the geodesic equation for single-metric gravity. As will be discussed below, for massless particles (such as photons, as the main connections to observations in cosmology), proportional metrics are of particular interest. The reason is that in this case the particles follow null geodesics with respect to both metrics and therefore one can safely use only one metric as the physical one. We will see in the following sections that this observation about proportional metrics is in fact very useful when considering the implications of the bimetric theory for cosmology.  There eq. (\ref{eq:DustOnlyHubble}) seems to provide this connection through the Hubble parameter $H$ as the quantity that appears in all standard observables in cosmology (see e.g.~\cite{Akrami:2012vf}).
Since our cosmological examples below feature this proportionality, there is no issue in choosing/deducing the ``physical'' metric.  

\section{The homogeneous and isotropic universe}
\label{thaib}

\subsection{Metrics and Bianchi constraints}

We now turn to the cosmologically interesting consequences of coupling both metrics to matter. We follow the standard recipe in cosmology and consider a homogeneous and isotropic background, for which the metrics are assumed to have the following Friedmann-Lema\^{\i}tre-Robertson-Walker (FLRW)-like forms~\cite{vonStrauss:2011mq,Akrami:2012vf}:
\begin{eqnarray}
  ds^2_g &=& - dt^2 + a^2\left(\frac{dr^2}{1 - kr^2} + r^2\left(d\theta^2 + \sin^2\theta d\phi^2\right)\right), \label{FRWg}\\
  ds^2_f &=& - X^2dt^2 + Y^2\left(\frac{dr^2}{1 - kr^2} + r^2\left(d\theta^2 + \sin^2\theta d\phi^2\right)\right), \label{FRWf}
\end{eqnarray}
where $a(t)$ and $Y(t)$ are the scale factors for spatial components of $g$ and $f$, respectively, and $X(t)$ is a similar function of time that corresponds to the temporal component of $f$. $k=\{-1,0,+1\}$ is the spatial curvature that is assumed to be the same for both metrics~\cite{vonStrauss:2011mq,Akrami:2012vf}).

Before moving on to the equations of motion for the gravity and matter components in this case, let us look at the full Bianchi constraint (\ref{eq:BianchiCombined}) which now becomes:
\begin{equation}
  \frac{3m^2M_g^2}{2}\left(1 - 1\right)\left(\beta_1 + 2\frac{Y}{a}\beta_2 + \frac{Y^2}{a^2}\beta_3\right)\left(\dot{Y} - \dot{a}X\right) = 0.
\end{equation}
As anticipated earlier (based on~\cite{HassanPrivate}), this equation is automatically satisfied and does not impose any conditions on the metric components $a$, $Y$ and $X$ (in contrast to the singly-coupled Hassan-Rosen bigravity~\cite{vonStrauss:2011mq,Akrami:2012vf}).

\subsection{Generalised field equations and matter equations of motion}

For the FLRW-like metrics (\ref{FRWg}) and (\ref{FRWf}), the generalised Einstein field equations (\ref{eq:Einsteingeng}) and (\ref{eq:Einsteingenf}) result in the following generalised Friedmann equations:
\scriptsize{
\begin{eqnarray}
  &&3\left(\frac{\dot{a}}{a}\right)^2  + 3\frac{k}{a^2} - m^2\left[\beta_0 +
3\beta_1\frac{Y}{a} + 3\beta_2\frac{Y^2}{a^2} + \beta_3\frac{Y^3}{a^3}\right] = 
-\frac{1}{M_g^2}T_{g0}^0, \label{eq:firstFriedmanng} \\
  &&-2\frac{\ddot{a}}{a}-\left(\frac{\dot{a}}{a}\right)^2 - \frac{k}{a^2}+
m^2\left[\beta_0 + \beta_1\left(2\frac{Y}{a} + X\right) +
\beta_2\left(\frac{Y^2}{a^2} + 2\frac{YX}{a}\right) + \beta_3
\frac{Y^2X}{a^2}\right] = \frac{1}{M_g^2}T_{g1}^1, \label{eq:spacialFieldEqg} \\
  &&3\left(\frac{\dot{Y}}{YX}\right)^2 + 3\frac{k}{Y^2} 
    - \frac{m^2}{M_\star^2}\left[\beta_4 + 3\beta_3\frac{a}{Y} +
    3\beta_2\frac{a^2}{Y^2} + \beta_1\frac{a^3}{Y^3}\right] 
    = -\frac{\alpha}{M_f^2}T_{f0}^0, \label{eq:firstFriedmannf} \\
  &&\frac{m^2}{M_\star^2}\left[\beta_4 + \beta_3\left(2\frac{a}{Y} +
      \frac{1}{X}\right) + \beta_2\left(\frac{a^2}{Y^2} +
      2\frac{a}{YX}\right) + \beta_1\frac{a^2}{Y^2X}\right] + 2\frac{\dot{Y}\dot{X}}{X^3Y} - 2\frac{\ddot{Y}}{YX^2} 
    - \left(\frac{\dot{Y}}{YX}\right)^2 - \frac{k}{Y^2} = \frac{\alpha}{M_f^2}T_{f1}^1, \nonumber \label{eq:spacialFieldEqf} \\
  \end{eqnarray}}\normalsize
where we have assumed $T_{g1}^1 = T_{g2}^2 = T_{g3}^3$ and $T_{f1}^1 = T_{f2}^2 = T_{f3}^3$, which is consistent with the symmetries of the metrics.

In order to analyse the above set of equations, we need to calculate the components of the stress-energy tensors $T_{g0}^0$, $T_{f0}^0$, $T_{g1}^1$ and $T_{f1}^1$ for cosmological fluids. We do this by using the actions of a set of $N$ point particles with masses $m_p^a (a=1,...,N)$ with respect to the two metrics $g$ and $f$ (here we follow the standard procedure given e.g. in~\cite{Peebles1993}:
\footnotesize{
\begin{equation}
  S_m^g = \sum_a m_p^a\int d^4x \delta\left(\mathbf{x} - \mathbf{x}_a(t)\right)\left(g_{\alpha\beta}\dot{x}^\alpha_a\dot{x}^\beta_{a}\right)^{\frac{1}{2}}, \qquad S_m^f = \sum_a m_p^a\int d^4x \delta\left(\mathbf{x} - \mathbf{x}_a(t)\right)\left(f_{\alpha\beta}\dot{x}^\alpha_a\dot{x}^\beta_{a}\right)^{\frac{1}{2}}.
\end{equation}}
\normalsize

The stress-energy tensors can now be obtained from the actions; for example for $T_g^{\mu\nu}$ this gives:
\begin{eqnarray}
  T_g^{\mu\nu} &=& \frac{2}{\sqrt{-\det g}}\frac{\partial}{\partial
g_{\mu\nu}}\sum_a m_p^a\delta\left(\mathbf{x} -
\mathbf{x}_a(t)\right)\left(g_{\alpha\beta}\dot{x}^\alpha_a\dot{x}^\beta_{a}\right)^{\frac{1}{2}}\nonumber\\
  &=& \sum_a m_p^a\frac{\delta\left(\mathbf{x} -
\mathbf{x}_a(t)\right)}{\sqrt{-\det g}}\frac{\dot{x}^{\mu}\dot{x}^\nu}{\left(g_{\alpha\beta}\dot{x}^{\alpha}\dot{x}^{\beta}\right)^{\frac{1}{2}}} \nonumber \\
&=& \sum_a m_p^a\frac{\delta\left(\mathbf{x} - \mathbf{x}_a(t)\right)}{\sqrt{-\det g}}\frac{\dot{x}^{\mu}\dot{x}^\nu}{\left(g_{00} + v^iv^jg_{ij}\right)^{\frac{1}{2}}},\label{eq:StressEnergyTensorFluid}
\end{eqnarray}
where $\dot{x}^\mu \equiv dx^{\mu}/dt$, $v^i \equiv dx^i/dt$ ($i = 1,...,3$), and the last expression is a consequence of the fact that in our case we have no space-time-mixing elements in $g_{\mu\nu}$. The expression for $T_f^{\mu\nu}$ can be derived in a similar way.

We should now investigate whether we can relate different components of the stress-energy tensors $T_g^{\mu\nu}$ and $T_f^{\mu\nu}$ in such a way that the set of field equations (\ref{eq:firstFriedmanng}), (\ref{eq:spacialFieldEqg}), (\ref{eq:firstFriedmannf}) and (\ref{eq:spacialFieldEqf}) can be solved with as few free parameters for the matter sector as possible. We ideally want to write the stress-energy components in one metric in terms of the ones in the other metric. This can be achieved by looking at the ratio of the corresponding components of the two tensors for our FLRW-like metrics (\ref{FRWg}) and (\ref{FRWf}):
\begin{eqnarray}\label{eq:StressEnergyRatio}
  \frac{T^{\mu\nu}_f}{T^{\mu\nu}_g} &=&
\alpha\frac{\sqrt{-\det g}}{\sqrt{-\det f}}\left(\frac{g_{\alpha\beta}\dot{x}^{\alpha}\dot{x}^{\beta}}{f_{\alpha\beta}\dot{x}^{\alpha}\dot{x}^{\beta}}\right)^{\frac{1}{2}} \nonumber \\
&=& \alpha\frac{a^3}{XY^3}\left(\frac{g_{\alpha\beta}\dot{x}^{\alpha}\dot{x}^{\beta}}{f_{\alpha\beta}\dot{x}^{\alpha}\dot{x}^{\beta}}\right)^{\frac{1}{2}} \nonumber \\
&=& \alpha\frac{a^3}{XY^3}\left(\frac{- 1 + a^2v^2}{-X^2+Y^2v^2}\right)^{\frac{1}{2}}, \label{eq:seratio}
\end{eqnarray}
where $v^2={v^1}^2+{v^2}^2+{v^3}^2$. For radiation and other relativistic particles $v=1$, and for pressureless particles (or dust) $v=0$.

A particularly interesting case is the relation that can be imposed on the energy densities of matter components, $\rho_g$ and $\rho_f$, by eq. (\ref{eq:seratio}). As for the single-metric gravity, we define these quantities as $\rho_g \equiv -T_{g0}^0 = -g_{0\mu}T_g^{0\mu}$ and $\rho_f \equiv -T_{f0}^0 = -f_{0\mu}T_f^{0\mu}$. For our particular choice of metrics these then mean that $\rho_g = T_g^{00}$ and $\rho_f = X^2 T_f^{00}$. Now we can use eq. (\ref{eq:seratio}) to find the relation between $\rho_g$ and $\rho_f$:
\begin{equation}\label{eq:rhoratios}
  \frac{\rho_f}{\rho_g} =  \alpha\frac{Xa^3}{Y^3}\left(\frac{- 1 +
a^2v^2}{-X^2+Y^2v^2}\right)^{\frac{1}{2}}.
\end{equation}

One may want to define pressures $P_g$ and $P_f$ in a similar, standard way, i.e. $P_g \equiv T_{gi}^i = g_{i\mu}T_g^{i\mu}$ and $P_f \equiv T_{fi}^i = f_{i\mu}T_f^{i\mu}$ (no sum implied over $i$). These definitions of pressures, however, turn out in general to result in non-trivial properties, for example if we impose the equation of state $P_g = \rho_g/3$ for radiation in the $g$ metric, a similar equation {\it cannot} be satisfied for the pressure and density, $P_f$ and $\rho_f$, in the $f$ metric; such a definition is not consistent with the combination of eq. (\ref{eq:StressEnergyRatio}), for spatial components of the stress-energy tensors, and eq. (\ref{eq:rhoratios}). However, the situation for pressureless particles is an exception; in fact, the equation of state in $g$, i.e. $P_g=0$, {\it does} imply a similar one in $f$, i.e. $P_f=0$. This comes from the observation that if the spatial components for the stress-energy tensor vanish for one of the metrics, they also vanish for the other.

We now assume, as in standard cosmology, that the universe mainly consists of non-relativistic matter (or dust) and relativistic particles (or radiation), with energy densities $\rho_m$ and $\rho_\gamma$, respectively. From eq. (\ref{eq:rhoratios}) we get:

\begin{equation}\label{eq:DensityRatioDust}
\frac{\rho^m_f}{\rho^m_g} =  \alpha\frac{a^3}{Y^3}, \qquad \frac{\rho_f^\gamma}{\rho_g^\gamma} =  \alpha\frac{a^3}{Y^3}\left(\frac{-1 + a^2}{-1 + \frac{Y^2}{X^2}}\right)^{\frac{1}{2}},
\end{equation}
where matter and radiation are characterised by $v=0$ and $v=1$, respectively. 

Using eq. (\ref{eq:DensityRatioDust}) and being aware of the subtlety in the definition of pressures in the bimetric framework, we can write the set of generalised Friedmann equations (\ref{eq:firstFriedmanng}), (\ref{eq:spacialFieldEqg}), (\ref{eq:firstFriedmannf}) and (\ref{eq:spacialFieldEqf}) as:
\scriptsize{
\begin{eqnarray}
&&3\left(\frac{\dot{a}}{a}\right)^2  + 3\frac{k}{a^2} - m^2\left[\beta_0 +
3\beta_1\frac{Y}{a} + 3\beta_2\frac{Y^2}{a^2} + \beta_3\frac{Y^3}{a^3}\right]= 
\frac{1}{M_g^2}\left(\rho_g^m + \rho_g^\gamma\right), \label{eq:firstFriedmanngDustRad} \\
&&-2\frac{\ddot{a}}{a}-\left(\frac{\dot{a}}{a}\right)^2 - \frac{k}{a^2}+
m^2\left[\beta_0 + \beta_1\left(2\frac{Y}{a} + X\right) +
\beta_2\left(\frac{Y^2}{a^2} + 2\frac{YX}{a}\right) + \beta_3
\frac{Y^2X}{a^2}\right] = \frac{1}{M_g^2} P_g^\gamma, \label{eq:spacialFieldEqgDustRad} \\
&&3\left(\frac{\dot{Y}}{YX}\right)^2 + 3\frac{k}{Y^2} -
\frac{m^2}{M_{\star}^2}\left[\beta_4 + 3\beta_3\frac{a}{Y} +
3\beta_2\frac{a^2}{Y^2} + \beta_1\frac{a^3}{Y^3}\right] =
\frac{\alpha}{M_f^2}\frac{a^3}{Y^3}\left(\rho_g^m + \left(\frac{-1 + a^2}{-1 +
\frac{Y^2}{X^2}}\right)^{\frac{1}{2}}\rho_g^\gamma\right), \label{eq:firstFriedmannfDustRad} \\
&&\frac{m^2}{M_\star^2}\left[\beta_4 + \beta_3\left(2\frac{a}{Y} +
\frac{1}{X}\right) + \beta_2\left(\frac{a^2}{Y^2} +
2\frac{a}{YX}\right) + \beta_1 \frac{a^2}{Y^2X}\right] + 2\frac{\dot{Y}\dot{X}}{X^3Y} -2\frac{\ddot{Y}}{YX^2}-\left(\frac{\dot{Y}}{YX}\right)^2  - \frac{k}{Y^2} = \frac{\alpha}{M_f^2} P_f^\gamma, \label{eq:spacialFieldEqfDustRad} \nonumber \\
\end{eqnarray}}\normalsize
where $P_g^\gamma$ and $P_f^\gamma$ are the ``pressure-like" quantities for radiation, and are related through eq. (\ref{eq:StressEnergyRatio}).

\section{Late-time universe and cosmic acceleration}

\label{luca}

\subsection{The case of dust only}

In the presence of radiation, eqs. (\ref{eq:firstFriedmanngDustRad}), (\ref{eq:spacialFieldEqgDustRad}), (\ref{eq:firstFriedmannfDustRad}) and (\ref{eq:spacialFieldEqfDustRad}) are exceedingly complicated. However, in this paper we are mainly interested in the implications of the theory for the late-time evolution of the universe. Here we therefore adhere to the standard assumption that the contribution of radiation to the dynamics of the universe at late times is negligible compared to that of non-relativistic matter, or dust. In such a dust-dominated universe, the pressure-like quantities $P_g^\gamma$ and $P_f^\gamma$, can also be justifiably dropped from the equations.

In order to simplify the equations even further, let us investigate what we can gain by studying the conservation equation (\ref{eq:ParticleEOM}) for a cosmological fluid. Using the general expression (\ref{eq:rhoratios}) we get:
\scriptsize{
\begin{eqnarray}
  0&=&a^3\left(\dot{\rho}_g + 3\frac{\dot{a}}{a}\left(\rho_g + P_g\right)\right) + \alpha XY^3\left(\dot{\rho}_f + 3\frac{\dot{Y}}{Y}\left(\rho_f +  P_f\right)\right)\nonumber\\
&=& \frac{d}{dt}\left(a^3\rho_g\right) + \alpha X\frac{d}{dt}\left(Y^3\rho_f\right) +  P_g\frac{d}{dt}\left(a^3\right) + \alpha X P_f\frac{d}{dt}\left(Y^3\right)\nonumber \\
&=& \left(1 + \alpha^2 X\left(\frac{-1 +  v^2a^2}{-1 +
v^2\frac{Y^2}{X^2}}\right)^{\frac{1}{2}}\right)\frac{d}{dt}\left(a^3\rho_g\right) +  P_g\frac{d}{dt}\left(a^3\right) + \alpha^2 Xa^3\rho_g\frac{d}{dt}\left(\left(\frac{-1 +  v^2a^2}{-1 +
v^2\frac{Y^2}{X^2}}\right)^{\frac{1}{2}}\right) + \alpha X P_f\frac{d}{dt}\left(Y^3\right). \label{eq:conservation}\nonumber \\
\end{eqnarray}}\normalsize

For dust, this equation can be significantly simplified to ($v=0$ and $P^m_g=P^m_f=0$):
\begin{equation}\label{eq:EOMDust1}
   \left(1 + \alpha^2 X\right)\frac{d}{dt}\left(a^3\rho^m_g\right) = 0,
\end{equation}
which implies that either $X$ must take the constant value $-1/\alpha^2$, or $d\left(a^3\rho^m_g\right)/dt=0$. Here we choose the second option mainly because $X=-1/\alpha^2$ diverges in the $\alpha=0$, i.e. in the limit where the second metric decouples from matter. Our assumption then implies that (together with eq. (\ref{eq:conservation})):
\begin{eqnarray}\label{eq:EOMDust2}
   &&\frac{d}{dt}\left(a^3\rho^m_g\right) = a^3\left(\dot{\rho}^m_g + 3\frac{\dot{a}}{a}\rho^m_g \right) = 0, \label{consdustg}\\
   &&\frac{d}{dt}\left(Y^3\rho^m_f\right) = Y^3\left(\dot{\rho}^m_f + 3\frac{\dot{Y}}{Y}\rho^m_f \right) = 0. \label{consdustf}
\end{eqnarray}

On the other hand, the individual Bianchi constraints (\ref{eq:Bianchig}) and (\ref{eq:Bianchif}) for our cosmological case and for a general fluid become:
\begin{eqnarray}
&&3\frac{m^2}{2}\left(\frac{\dot{Y}}{a} - \frac{\dot{a}}{a}X\right)\left(\beta_1 + 2\beta_2\frac{Y}{a} + \beta_3\frac{Y^2}{a^2}\right) = -\frac{1}{M_g^2} \left(\dot{\rho}_g + 3\frac{\dot{a}}{a}\left(\rho_g + P_g\right)\right),\label{eq:BianchigHom}\\
&&3\frac{m^2}{2}\frac{a^3}{XY^3}\left(\frac{\dot{a}}{a}X - \frac{\dot{Y}}{a}\right)\left(\beta_1 + 2\beta_2\frac{Y}{a} + \beta_3\frac{Y^2}{a^2}\right) = -\frac{\alpha}{M_f^2}\left(\dot{\rho}_f + 3\frac{\dot{Y}}{Y}\left(\rho_f + P_f\right)\right).\label{eq:BianchifHom}
\end{eqnarray} 

Combining these with eqs. (\ref{consdustg}) and (\ref{consdustf}) for the dust-dominated universe then implies that:
\begin{equation}
  X = \frac{\dot{Y}}{\dot{a}}.
\end{equation}

This is the condition that holds in the original Hassan-Rosen theory where Bianchi constraints (\ref{eq:Bianchig}) and (\ref{eq:Bianchif}) are individually satisfied~\cite{vonStrauss:2011mq,Akrami:2012vf}. Therefore for the late-time universe, dominated by dust, the evolution follows the dynamical equations that resemble those in the singly-coupled case. It can be shown~\cite{vonStrauss:2011mq,Akrami:2012vf} that the generalised Friedmann equations in this case boil down to a set of two simple equations that fully determine the background evolution of the universe and can be written in terms of only two dynamical variables, namely the usual Hubble parameter $H = \dot{a}/a$ corresponding to the metric $g$, and $y \equiv Y/a$, the ratio of the spatial scale factors corresponding to the two metrics. These two equations are (assuming a flat universe $k=0$):
\small{
\begin{eqnarray}
  \frac{H^2}{H_0^2} &=& \frac{B_0}{3} + B_1y + B_2y^2 + \frac{B_3}{3}y^3 + \Omega_m,\label{eq:DustOnlyHubble}\\
  0 &=& \frac{B_3}{3}y^4 + \left(B_2 - \frac{B_4}{3}\right)y^3 + \left(B_1 - B_3\right)y^2 + \left(\Omega_m + \frac{B_0}{3} - B_2\right)y - \frac{B_1}{3} - \alpha\Omega_m, \label{eq:DustOnlyQuartic}
\end{eqnarray}}\normalsize
where $H_0$ is the value of $H$ at present time, $\Omega_m\equiv \rho_g^m/(3H_0^2M_g^2)$ and $B_n \equiv m^2\beta_n/H_0^2$. In addition, here we have performed the constant rescaling of parameters and metric components as described in \cite{Berg:2012kn,Akrami:2012vf}, to set $M_\star^2$ to unity. This can be done in this particular case of the dust-dominated universe because the matter Lagrangian has a very simple structure; clearly the coupling parameter $\alpha$ must also be rescaled accordingly.

\subsection{Emergent cosmological constant and $\Lambda$CDM regime}

It has already been shown in~\cite{Akrami:2012vf} that the singly-coupled (Hassan-Rosen) bigravity theory can yield good fits to various cosmological data (at least at the background level), even in the absence of an explicit cosmological constant (or vacuum energy) which is parameterised by $B_0/3$. As far as the evolution equations are concerned, our doubly-coupled ``bimetric" gravity closely resembles the Hassan-Rosen theory, with only one additional parameter $\alpha$. It is therefore natural to expect that the doubly-coupled bigravity also gives good fits to the data as its subset does.

We however know that the standard $\Lambda$CDM model of cosmology is {\it not} a subset of the Hassan-Rosen theory unless we set $\Omega_\Lambda = B_0/3$~\cite{Akrami:2012vf}. A novel feature of the doubly-coupled generalisation that we propose here is that ``it can yield exact $\Lambda$CDM even in the absence of an explicit cosmological constant (or vacuum energy), i.e. with $B_0/3=0$". In order to see this, let us look at the asymptotic behaviour of the theory at early and late times.

At early times, unless $y$ becomes unnaturally large in the past, the $\Omega_m$ terms dominate since $\Omega_m$ scales with redshift as $\Omega_m(z) = \Omega_m^0(1+z)^3$. Eq. (\ref{eq:DustOnlyQuartic}) then implies that at this limit $y \rightarrow\alpha$, and it can be seen from eq. (\ref{eq:DustOnlyHubble}) that in this case the evolution of the universe resembles that of $\Lambda$CDM:
\begin{equation}
  \Omega_\Lambda = \frac{B_0}{3}+B_1\alpha+B_2\alpha^2+\frac{B_3}{3}\alpha^3,\label{lambdaeff}
\end{equation}
even if we set $B_0/3=0$.

At late times on the other hand, the $\Omega_m$ terms die off and $y$, far in the future, will be given by generally a different constant value which is a solution to the quartic equation:
\begin{equation}\label{eq:LateTimeQuartic}
   \frac{B_3}{3}y^4 + \left(B_2 - \frac{B_4}{3}\right)y^3 + \left(B_1 - B_3\right)y^2 + \left(\frac{B_0}{3} - B_2\right)y - \frac{B_1}{3} = 0.
 \end{equation}

Therefore, the dynamical variable $y$ must transition from one early-time constant value to another late-time value over the history of the universe.

In addition, it can be proven that eq. (\ref{eq:DustOnlyQuartic}) has no local minima or maxima during the cosmic evolution. We realise this by taking the derivative of the equation with respect to the cosmic time and looking for cases where $\dot{y}\equiv dy/dt=0$. This requires:
\begin{equation}
  y|_{\dot{y}=0} = \alpha,
\end{equation}
which is the same as its early-time value. This means that if the function $y$ has decreased (increased) from its early-time value, then it must have increased (decreased) again to obtain this value before reaching the local maximum (minimum). Therefore there must have been a point at which $\dot{y}$ has become zero before the local maximum (minimum) point where it has again become zero. $y$ at this other local maximum (minimum) has to have had a value different from $\alpha$ which is not possible according to the evolution equation. This therefore proves that $y$ has to have been monotonic in time.

An immediate and very interesting consequence of this behaviour of $y$ is that if we choose $\alpha$ such that $y = \alpha$ is a solution of eq. (\ref{eq:LateTimeQuartic}), then $y$ must have held this constant value throughout its evolution. A constant $y$ then gives rise to an emergent cosmological constant given by eq. (\ref{lambdaeff}). By tuning $\alpha$, as well as the parameters $B_n$, this therefore shows that within the framework of doubly-coupled bimetric gravity, contrary to the singly-coupled case, it {\it is} always possible to produce an exact $\Lambda$CDM universe (at least at the background level), even in the absence of vacuum energy contributions, i.e. even with $B_0/3=0$.

Another interesting feature of this particular solution (constant $y$) is that now the two metrics of the theory are proportional:
\begin{equation}
  f_{\mu\nu} = y^2 g_{\mu\nu} = \alpha^2 g_{\mu\nu},
\end{equation}
which, as we discussed in section \ref{physical}, removes the difficulty in determining the cosmological observables in the theory. 

\section{Special parameter regimes}

\label{spr}

In this section, we further demonstrate the ability of the doubly-coupled bimetric gravity in mimicking the $\Lambda$CDM cosmology and producing an accelerated universe by studying interesting sub-models with special parameter configurations that exhibit novel features of the theory peculiar to the $\alpha \neq 0$ case.

\subsection{One non-zero $B$ parameter}

In addition to the study of the full, singly-coupled (Hassan-Rosen) bigravity model, the statistical analysis performed in \cite{Akrami:2012vf} included the analysis of, amongst others, three sub-models where all $B$-parameters except one of the parameters $B_1$, $B_2$ or $B_3$ were set to zero. In that exploration it was shown that the first of these cases (i.e. the only-$B_1$ case) could yield very good fits to the cosmological observations, whereas the other two cases (i.e. the only-$B_2$ and only-$B_3$ cases) could not. Here, we would therefore like to extend those analyses to the cases where the $B$-parameters are combined with a non-zero value for $\alpha$ to see whether this extension can result in better fits in the case of only-$B_1$ and/or relatively good fits in the cases of only-$B_2$ and only-$B_3$ models. In the following subsections we show analytically how the solutions of the theory in these cases can mimic an exact cosmological constant based on our discussions in the previous section on the monotonicity of the dynamical variable $y$ with time. In each case we give the exact value of the solution, as well as the coupling constant $\alpha$, in order for the model to yield the desired properties. We also present the values of the cosmological constant given by the models in terms of the model parameters.

\subsubsection{$B_1 \neq 0$ and $\alpha \neq 0$}
In this case the quartic equation (\ref{eq:DustOnlyQuartic}) reduces to a quadratic equation with the following solutions:
\begin{equation}
  y = \frac{-\Omega_m \pm \sqrt{\Omega_m^2 + 4B_1\left(\frac{B_1}{3} + \alpha\Omega_m\right)}}{2B_1}.
\end{equation}

It is then easy to see that if we choose $\alpha = 1/\sqrt{3}$, which is the late-time solution for $y$, then $y = \alpha = 1/\sqrt{3}$ is the positive-branch solution for $y$ for all times, i.e. $y$ becomes constant in time and therefore independent of $\Omega_m$ (which is the only quantity that varies with time). In this case, the model gives an exact cosmological constant with a value that is set only by $B_1$:
\begin{equation}
\Omega_\Lambda = \frac{B_1}{\sqrt{3}}.
\end{equation}

The model is therefore equivalent to the $\Lambda$CDM model which is known to give a perfect fit to the observational data.

\subsubsection{$B_2 \neq 0$ and $\alpha \neq 0$}
With only non-zero $B_2$ and $\alpha$, eq. (\ref{eq:DustOnlyQuartic}) becomes a cubic equation. Considering instead eq. (\ref{eq:LateTimeQuartic}), we realise that the positive, non-trivial, late-time solution for $y$ is $y = 1$. By setting $\alpha = 1$ in eq. (\ref{eq:DustOnlyQuartic}), we observe that $y = 1$ is a positive-branch solution for this equation at all times. The cosmological constant contribution, given by the $B_2 y^2$ term, will then be simply:
\begin{equation}
\Omega_\Lambda = B_2,
\end{equation}
which shows that this sub-model is now equivalent to the $\Lambda$CDM model of cosmology and therefore is able to explain the observations as well as the $\Lambda$CDM does (in contrast to the singly-coupled case).

Since $y$ enters only quadratically in the Friedmann equation (\ref{eq:DustOnlyHubble}), a choice of $y = \alpha = -1$ would yield the same results here. However, regarding the definition of $y$ as the ratio of the two spatial scale factors $Y$ and $a$, corresponding to the metrics $f$ and $g$, a negative value for it may sound unphysical and we therefore prefer the positive solution, $y=1$.  Also a choice of negative $\alpha$ leads to negative energy terms in the $f$-sector which might lead to ghosts or other pathologies.

In the case of $y = \alpha = 1$, we can also note that the theory becomes completely symmetric in terms of the metrics as both metrics couple to matter with the same coupling strengths.

\subsubsection{$B_3 \neq 0$ and $\alpha \neq 0$}
In the case with only $B_3$ and $\alpha$ non-zero we find that the positive non-trivial solution to the late-time eq. (\ref{eq:LateTimeQuartic}) is $y = \sqrt{3}$. Inserting $\alpha = \sqrt{3}$ into eq. (\ref{eq:DustOnlyQuartic}) we see that $y = \sqrt{3}$ is the positive solution at all times. This means that the cosmological constant contribution in this case, which is given by $(B_3/3)y^3$ in the Friedmann equation (\ref{eq:DustOnlyHubble}), becomes:
\begin{equation}
\Omega_\Lambda = \sqrt{3}B_3.
\end{equation}

Again, this shows that the only-$B_3$(and $\alpha$)-nonzero subset of the theory {\it does} provide a good fit to the data, a feature that is new to the doubly-coupled compared to the singly-coupled bigravity theory.

Similar to the case of only $B_2$ and $\alpha$ nonzero studied above, here again a negative solution ($y = \alpha = -\sqrt{3}$) is also allowed by the equations. However, in this case in addition to the arguments stated in the previous case for rejecting such a solution, the cosmological constant would be negative, which is inconsistent with observations.

\subsection{The partially massless case}
In \cite{Hassan:2012gz}, a partially massless (PM), singly-coupled, bimetric gravity theory is introduced based on the Hassan-Rosen theory, where (setting $M_\star = 1$):
\begin{equation}
  B_0 = B_4 = 3B_2, \qquad B_1 = B_3 = 0. \label{PMparams}
\end{equation}

With these particular relations between the parameters of the theory, the theory is at the so-called Higuchi bound~\cite{Higuchi:1986py,Higuchi:1989gz}, and on a pure de Sitter background the mass of the spin-2 state (the so-called Fierz-Pauli (FP) mass $m_{FP}$) can be fully determined in terms of the cosmological constant $\Lambda$ through the following relation:
\begin{equation}
  m_{FP}^2=\frac{2}{3}\Lambda.
\end{equation}

The PM theory is very interesting mainly because a new gauge symmetry appears at the Higuchi bound which decouples the potentially dangerous, helicity-zero component of the spin-2 field, and leaves only four healthy propagating modes.

However, in the context of the singly-coupled bigravity theory, i.e. when $\alpha = 0$, one can see that (in the absence of a curvature term) this theory gives no non-GR dynamics. It can of course still include a traditional cosmological constant given by $B_0/3$, but this case is relatively uninteresting as the gravity sector will reduce to the standard (massless) gravity where partial masslessness no longer makes sense. However, as we demonstrate here, in the doubly-coupled theory, this picture completely changes.

In order to see this, let us study the consequences of the quartic equation (\ref{eq:DustOnlyQuartic}) for the PM case. On a pure de Sitter background, i.e. where $\Omega_m=0$, by imposing the relations (\ref{PMparams}) between the model parameters, eq. (\ref{eq:DustOnlyQuartic}) is clearly satisfied independently of the value of $y$. This is a manifestation of the new gauge symmetry introduced by the PM theory. We, however, know that we do not live in a pure de Sitter universe and we therefore need to include the $\Omega_m$ terms in eq. (\ref{eq:DustOnlyQuartic}). In this case, the PM conditions on $B_n$ parameters (\ref{PMparams}) imply that:
\begin{equation}
  \Omega_m y - \alpha \Omega_m = 0 \Rightarrow y = \alpha.
\end{equation}

In the original Hassan-Rosen theory, this means $y=0$, i.e. we are back to GR. However, in the doubly-coupled regime, with $\alpha \neq 0$, even a flat universe can display non-trivial solutions. Eq. (\ref{eq:DustOnlyHubble}) then becomes (since $y=\alpha$):
\begin{equation}\label{partiallyMasslessSolution}
\frac{H^2}{H_0^2} = B_2\left(1 + \alpha^2\right) + \Omega_m.
\end{equation}

Therefore, in this PM doubly-coupled case the contribution to the energy density from the bigravity sector is {\it purely} a cosmological constant since the solution for $y$ is completely independent of $\Omega_m$ at all times. This means that this theory can fit the cosmological observations (at the background level) just as well as the standard $\Lambda$CDM model.

In addition, we note from eq. (\ref{partiallyMasslessSolution}) that the first contribution to $\Omega_\Lambda$, i.e. $B_2$, comes from the explicit cosmological constant $B_0/3$. It is, interestingly, possible here to start with an arbitrarily small such term and magnify its effect through the $\alpha$ parameter to obtain a suitable value for the full cosmological constant. In summary, good fits to observations for this theory correspond to values of $B_2$ and $\alpha$ where $B_2\left(1 + \alpha^2\right) = \Omega_\Lambda^O$, and $\Omega_\Lambda^O$ is the best-fit value for the cosmological constant density parameter in the $\Lambda$CDM model.

Finally, it is important to note another connection between PM and doubly-coupled gravities that makes it very interesting to study the two theories in a unified framework. The PM parameter setup treats the two metrics on an equal footing, and introduces a symmetry between them. Coupling only one metric to matter breaks this symmetry strongly, whereas coupling both metrics seems to give a somewhat softer breaking of the symmetry. Therefore, studying the PM gravity in the context of the doubly-coupled bimetric gravity is a natural procedure. In addition, if the bigravity theory is a correct description of nature, one could consider the fact that a cosmological constant is a good explanation of dark energy at the background level as an indication of the symmetry of the PM theory being present and only very softly broken. In the PM case a gauge symmetry is in place to insure that the theory is only governed by a cosmological constant, and that the two metrics must be proportional.

\subsection{Remarks on special parameter scenarios}

Based on our results and discussions on the cosmological scenarios for the late-time evolution of the universe in the special parameter regimes described above, here we end the present section by a couple of interesting remarks.

First, we saw that it has been possible to tune the coupling constant $\alpha$ in the cases of ``one non-zero $B$ parameter" models in such a way that the theory reduces to the pure $\Lambda$CDM model where the value of $\Lambda$ (or $\Omega_\Lambda$) is determined purely by the value of the non-zero $B$ parameter present in each case. However, it is clearly possible in those cases that the parameters do not ``conspire" perfectly to give an exact $\Lambda$CDM cosmology. It might be the case that the values of the parameters are such that the resultant model resembles the $\Lambda$CDM dynamics to a great extent (which must be the case since we know that $\Lambda$CDM is a good phenomenological description of the universe), but still slightly deviates from it. In this case one can have a model of dynamical dark energy (since $y$ can change with time) which while being indistinguishable from the $\Lambda$CDM model at the background level, can be distinguished for example by perturbative analysis. Therefore it is not unnatural to perform a statistical analysis of each model, similar to the work that has been done in~\cite{Akrami:2012vf} but using more data at both background and perturbative levels. This way one will be able to test whether the best-fit parameters prefer a pure cosmological constant  or a varying dark energy. We leave the investigation of these cases for future work.\footnote{Before performing such an analysis it is necessary to clarify how to connect the theory to observables in the generic case where $g$ and $f$ are not proportional. As the theory is close to $\Lambda$CDM, though, a perturbative approximation to this problem might be applicable.}

Our second remark concerns the observation that, in contrast to the ``one non-zero $B$ parameter" models, the PM scenario in the doubly-coupled bimetric framework admits only pure $\Lambda$CDM cosmology, i.e a cosmological constant interpretation of dark energy. 
This is interesting as the proportionality of the two metrics in this case strengthens the symmetry between the two: the null geodesics defined by either of the metrics coincide.  As we saw before, this simplifies the bridge between observational quantities and theory and makes unambiguous geometric definitions of massless particles possible.

\section{Conclusions and perspectives}
\label{c}

In this paper we have introduced a truly bimetric theory of gravity based on the ghost-free Hassan-Rosen theory first presented in \cite{Hassan_et_al2012a,Hassan_et_Rosen2012a,Hassan:2011ea} having now both of the metrics coupled to matter in a symmetric way. While this theory retains many of the attractive features of the original theory, its dynamical solutions are in general different (except in vacuum of course). The profound difference lies in the new coupling  and the possibilities they entail for new experimental tests of the possible bimetric nature of space-time. 

The particular coupling between the metrics inherited from the Hassan-Rosen theory guarantees that the system is not overly constrained as is the case in a bimetric theory which features two copies of GR. Instead the equations of motion change to include terms dependent on both metrics, as we have shown in the particular cases of a scalar field and a point particle. The coupling parameter $\alpha$ for the second metric can be adjusted to avoid immediate conflicts with current gravitational experiments; the experimental signatures in solar system and astrophysical tests of gravity would be a worthwhile subject for further work.

Here, as the first concrete example, we considered the implications of this theory to flat, late-time cosmology. In this context we saw that the theory became considerably simplified due to the symmetries of the cosmological solutions and in particular special properties of solely non-relativistic matter. In the late-time universe where only dust-like sources are relevant, the background dynamics of the doubly-coupled theory bears strong resemblance to that of the original Hassan-Rosen theory as discussed in \cite{vonStrauss:2011mq,Akrami:2012vf}. However, whereas the Hassan-Rosen theory could only give an exact cosmological constant evolution in the case where the theory in fact only featured an exact cosmological constant, i.e. only $B_0$ nonzero, we showed that the doubly-coupled version could yield exact cosmological constant dynamics for an arbitrary choice of the parameters $B_n$ with an appropriate choice of the parameter $\alpha$. In particular we gave exact solutions for this in the cases of a single non-zero parameter $B_1$, $B_2$ or $B_3$. It will be of interest to study the formation of structure in these models.

Another emergent property of the doubly-coupled bimetric theory for cosmology is that the partially massless parameter combination, rather than giving just a trivial solution for the $f$-metric, gives a cosmological constant theory where $f$ is proportional to $g$. An interesting feature of such solutions is that they render the two metrics equal in defining the physical observables making the connection between theory and observations obvious. In general however, this seems to pose an intriguing fundamental issue in the new bimetric theory.

\begin{acknowledgments}
We thank Tessa Baker, Iain A. Brown, Philip Bull, Jonas Enander, Pedro G. Ferreira, S. F. Hassan, Claudio Llinares, Edvard M\"{o}rtsell, Martin Sahl\'{e}n, Angnis Schmidt-May, Mikael von Strauss, Bo Sundborg and Hans A. Winther for enlightening and helpful discussions. We are particularly grateful to S. F. Hassan for sharing valuable remarks on theoretical foundations of the present work, and Pedro G. Ferreira, Timothy Clifton, Johannes Noller and James H. C. Scargill for useful comments on the manuscript. Y.A. is supported by the European Research Council (ERC) Starting Grant StG2010-257080 and T.S.K. is supported by the Research Council of Norway. D.F.M. thanks the Research Council of Norway FRINAT grant 197251/V30. D.F.M. is also partially supported by project CERN/FP/123618/2011 and CERN/FP/123615/2011.
\end{acknowledgments}

\bibliographystyle{JHEP}
\bibliography{sources}

\end{document}